\pdfoutput=1
\documentclass[aip,reprint,citeautoscript,showpacs]{revtex4-1}

\usepackage{graphicx}
\usepackage[utf8]{inputenc}

\newcommand{\un}[1]{{\ensuremath{\,\text{#1}}}}
\newcommand{\PdNi}{\ensuremath{\text{Pd}_{0.3} \text{Ni}_{0.7}}}

\usepackage{color}

\begin{document}

\title{Transversal Magnetic Anisotropy in Nanoscale PdNi-Strips}

\author{D. Steininger}
\email{daniel.steininger@physik.uni-r.de}
\author{A. K. Hüttel}
\author{M. Ziola}
\author{M. Kiessling}
\author{M. Sperl}
\author{G. Bayreuther}
\author{Ch. Strunk}
\affiliation{Institute for Experimental and Applied Physics, University of
Regensburg, Universitätsstr.\ 31, D-93053 Regensburg, Germany}

\begin{abstract}
We investigate submicron ferromagnetic PdNi thin-film strips intended as contact
electrodes for carbon nanotube-based spintronic devices. The magnetic
anisotropy and micromagnetic structure are measured as function of temperature
and aspect ratio. Contrary to the expectation from shape anisotropy, magnetic
hysteresis measurements of \PdNi\ on arrays containing strips of various width
point towards a magnetically easy axis in the sample plane, but transversal to
the strip direction. Anisotropic magnetoresistance measured on individual
\PdNi\ contact strips and magnetic force microscopy images substantiate that conclusion.
\end{abstract}


\maketitle

%=====================================================================

\section{Introduction}
Carbon nanotubes (CNTs) have become a frequently used material for spin
dependent transport experiments within the past years. Due to spin
lifetimes of several nanoseconds \cite{alphenaar, Hueso}, CNTs provide excellent
conditions for this kind of investigation \cite{Kontos2005, preusche,
splitkondo, aurich}. Spin injection
requires ferromagnetic contact electrodes to the CNT which provide both spin
polarization and a transparent electric contact. As palladium is known for its
high contact transparency to CNTs \cite{biercuk, Kontos2005, Sahoo2005,
palma, splitkondo}, an alloy of palladium with nickel as ferromagnetic material
was chosen for this study. The electronic properties of pure palladium
\cite{mueller, hodges} and especially the formation of giant magnetic moments in
dilute alloys with small amounts of ferromagnetic elements have been well
studied for
both bulk material and thin films, with respect to Curie points and critical
alloy concentrations for the onset of ferromagnetism \cite{ododo, ododo2,
nieuwenhuys, wollan, loram, arham}.

Objective of this work is the investigation of thin film \PdNi\ high aspect
ratio shapes, particularly with regard to to their suitability as ferromagnetic
contact electrodes for CNT devices. While the similar alloy
$\text{Pd}_{1-x} \text{Fe}_{x}$ is governed by shape anisotropy \cite{preusche,
silvia, ododo, ododo2, ododo3}, this finding is not transferable to
$\text{Pd}_{1-x} \text{Ni}_{x}$ due to
the significantly different crystal structures of nickel and iron and the
differring ferromagnet concentrations in the experiments.\cite{chauleau}

We use \PdNi\ strips at widths of $250\un{nm}$ --- $1500\un{nm}$, maintaining a
constant length of $5\,\mu\text{m}$ and metal film thickness of $50\un{nm}$. All
samples were fabricated on a boron doped, p++ type $\text{Si}$ substrate with
thermally grown $\text{SiO}_2$ surface oxide, and structured via electron beam
lithography and lift-off process using a polymethyl methacrylate two layer
resist. The deposition of the thin films was done by electron beam evaporation
from pre-mixed bulk material in an UHV evaporation system with a base pressure
of $4 \times 10^{-8}\un{mbar}$.
\begin{figure}[th]
\includegraphics[width=6cm]{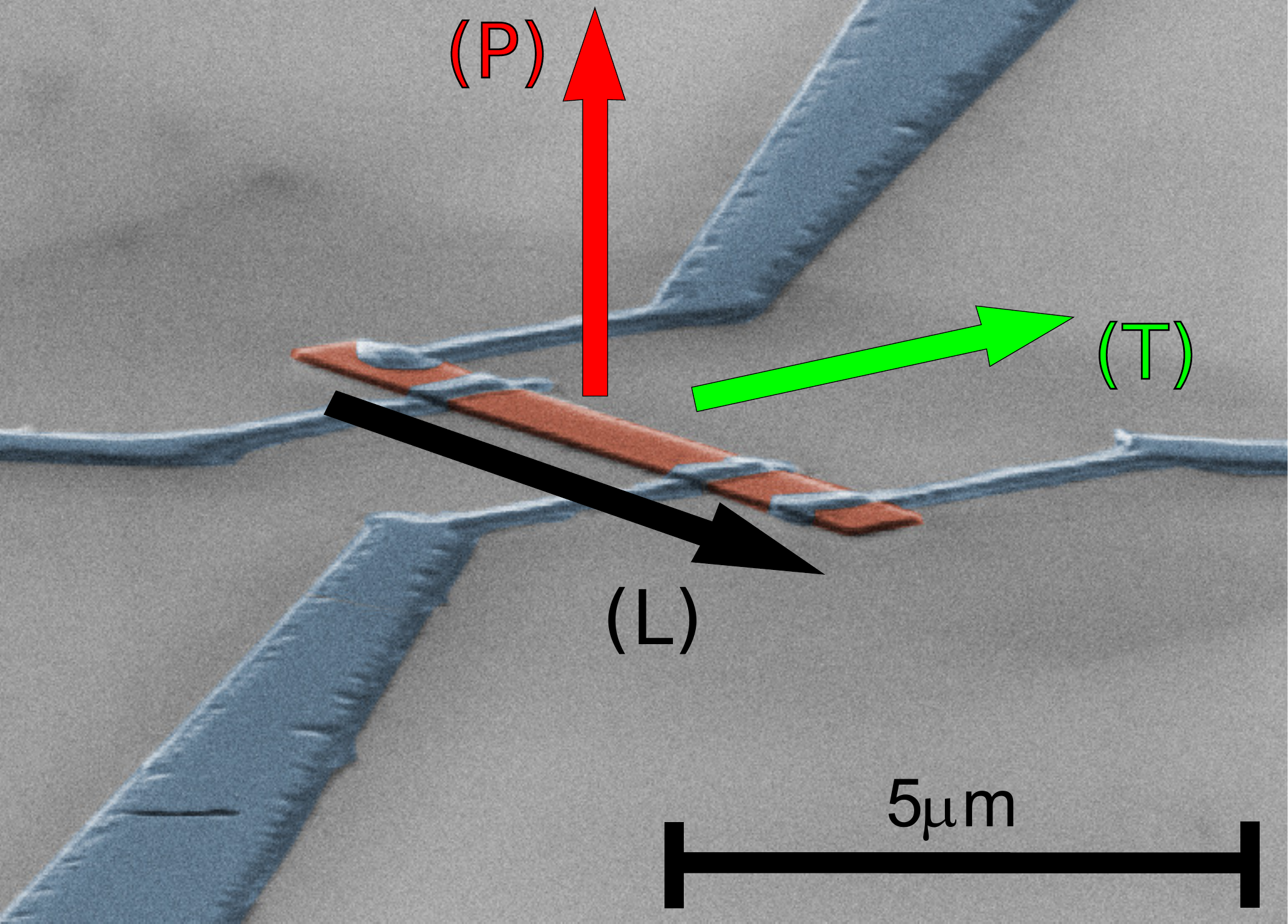} 
\caption{(Color online) Scanning electron microscopy (SEM) image of a
contacted \PdNi\ strip (colorized in red). The four contact leads
(colorized blue) were evaporated in a second step from pure palladium. They
provide the possibility to perform four terminal resistance measurements. The
arrows indicate the three tested orientations of the external magnetic field
relative to the strip, namely black (L): longitudinal, red (P): perpendicular,
and green (T): transverse.}
\label{rem}
\end{figure}
Figure \ref{rem} displays such a strip pattern, in this particular case
with leads attached in a second lithography step for magnetoresistance
measurements. It also introduces the characteristic directions for all the
magnetic field orientations used in this manuscript, i.e., perpendicular (P),
transverse (T), and longitudinal (L) with respect to the $\text{Pd}_{0.3}
\text{Ni}_{0.7}$ strip.

\section{SQUID magnetization measurements}
Superconducting quantum interference device (SQUID) magnetization measurements
using a commercial magnetometer have been used to characterize the
averaged magnetic hysteresis of arrays containing $1.3 \times 10^6$
non-contacted, electrically isolated thin film strips. They reveal the switching characteristics at various temperatures and orientations of the external magnetic field. 
\begin{figure}[th]
\begin{center}
\includegraphics[width=8.5cm]{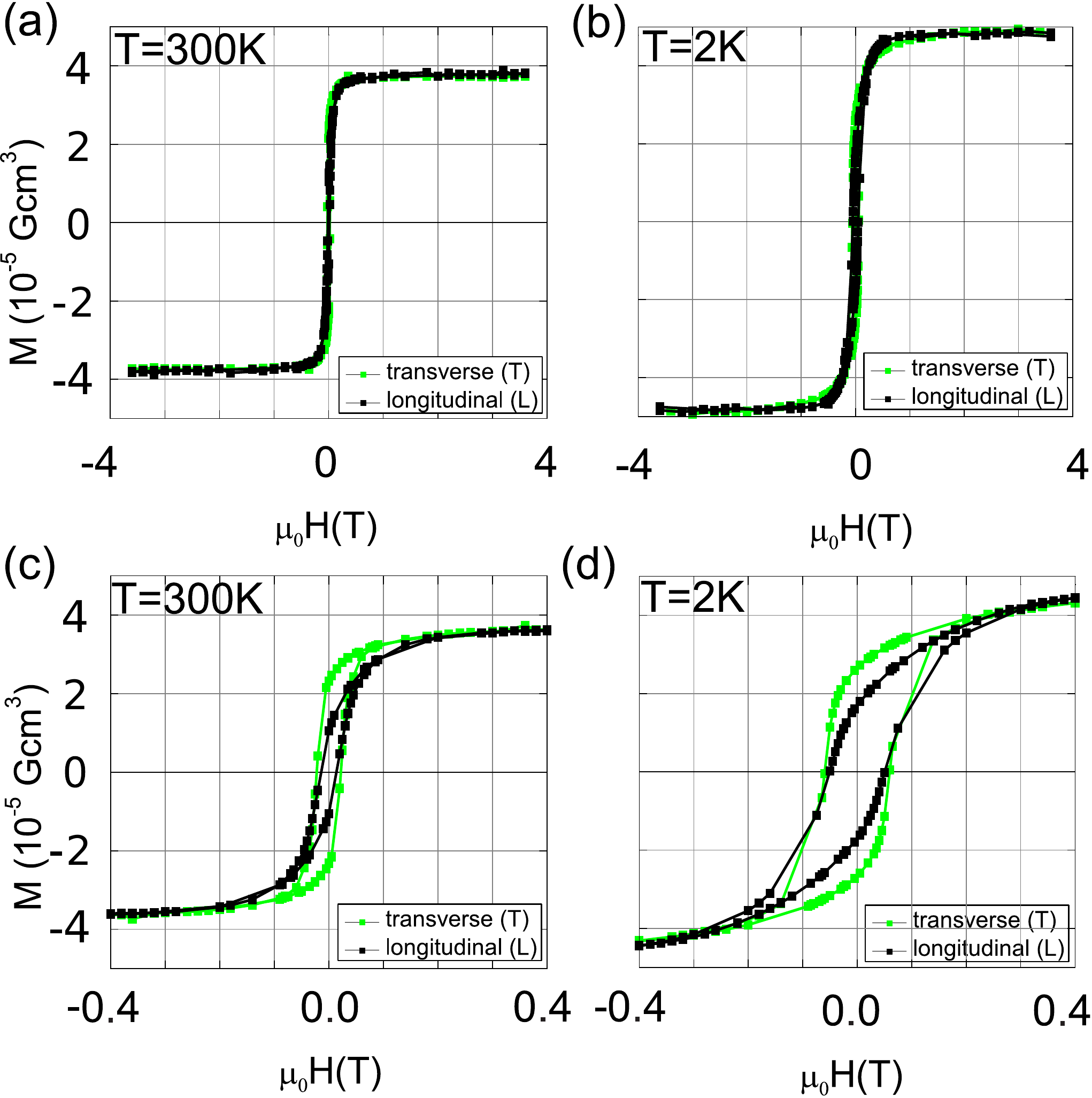}
\end{center}
\caption{(Color online) SQUID magnetization measurements on an array containing
$1.3 \times 10^6$ strips of $250\un{nm}$ width and $5\,\mu\text{m}$ length.
The array was measured in an external field of $\mu_{0}\left|H\right| \leq
3.6\un{T}$ in both transverse (T,
green) and longitudinal (L, black) direction (see also Fig.~\ref{rem}) at
$T=300\un{K}$ (a,c) and $T=2\un{K}$ (b,d). The upper panels (a,b) show the full
field range of the hysteresis loop while the lower panels (c,d) zoom into the
range $-0.4\un{T} \leq \mu_{0}H \leq 0.4\un{T}$.} 
\label{squid}
\end{figure}
Fig.~\ref{squid} displays the magnetization curves of an array of $250\un{nm}$
wide strips for two different temperatures after subtraction of the diamagnetic and paramagnetic moments of the boron doped Si substrate, the latter showing a strongly field-dependent susceptibility at low temperature which was fitted with a Brillouin function.
The total magnetic moment of the sample at T = 300K is $M_{\text{tot}} = 3.7 \times 10^{-5}\un{Gcm}^3$ (see Fig.~\ref{squid}(a)), from which an average magnetic moment of $\overline{\mu} = 0.583 \mu_B$ per alloy atom can be calculated. This value is comparable to previous results
for bulk \PdNi, where also the Curie temperature of this material
has been characterized (\un{$T_C$} $\approx 532$K).\cite{ododo,ododo2}
Figures~\ref{squid}(a-b) show magnetization loops in a transverse (T) and longitudinal (L) external field at $T=300\un{K}$
and $T=2\un{K}$, while Figures~\ref{squid}(c-d) zoom in onto the range
$-0.4\un{T} \leq \mu_{0} H \leq 0.4\un{T}$.
The higher field required for saturation and the larger coercive field values observed at low temperature can be explained by the absence of thermally activated motion of domain walls. Saturation is reached at lower external fields for the transverse orientation of external magnetic field. Also the remanent magnetization in transverse direction is significantly higher than in longitudinal orientation. These observations indicate an initially unexpected magnetic anisotropy with the easy axis along the transverse direction (T).

Concerning the details of the magnetization curves, the remanent magnetization in (T) direction is not 100\% as expected for an easy axis. This can be explained by the strong demagnetizing field connected with the magnetization pointing perpendicular to the edge of the sample. This is expected to cause some reversed domains as indicated by MFM images discussed below (Fig. \ref{mfm}(b)). Furthermore, to avoid a large related magnetostatic energy the magnetization will rotate towards the (L) direction close to the lateral edges. Hence, the resulting average transverse magnetization component decreases with decreasing strip width and becomes substantially reduced for sub-micrometer dimensions.
On the other hand, the remanence in the (L) direction is not zero as would be expected for a uniaxial hard axis. This is presumably a result of the domain splitting in the remanent state after saturation along the hard axis, which is confirmed by the MFM images shown in Fig. \ref{mfm}(a) (see below). For small domains with a width below 300 nm as deduced from Fig. \ref{mfm}(a) the magnetization inside and in the vicinity of the Néel type domain walls will have a substantial component perpendicular to the wall, i.e. along the (L) direction, which shows up as the longitudinal remanence.

\section{Magnetotransport}

To obtain further information on the direction of spontaneous magnetization
and the switching behaviour of single contact electrodes, the anisotropic
magnetoresistance (AMR) of individually contacted \PdNi\ strips was measured. As
with the samples for the SQUID measurements, the strips were fabricated via
electron beam lithography and lift-off process on identical $\text{Si} /
\text{SiO}_2$ substrate material. The non-magnetic palladium leads (see
Fig.~\ref{rem}) were deposited using a second electron beam lithography and
evaporation step. The resistance of a ferromagnetic strip is generally higher
when the
current through the strip is aligned parallel to the magnetization vector of the
structure ($\vec{\text{J}}\parallel\vec{\text{M}}$) than in the perpendicular
case ($\vec{\text{J}}\perp\vec{\text{M}}$).\cite{footnote1}$^,$\cite{mcguire,
gonzalez, senoussi} Transport measurements were performed at
$T=4.2\un{K}$, applying a maximum external magnetic field of $\mu_0
H_{\text{max}}$=$\pm 925\un{mT}$ successively in different orientations relative
to the strip. The resistance measurement itself was carried out using a
four-terminal setup and lock-in technique.

\begin{figure}[th]
\begin{center}
\includegraphics[width=8.5cm]{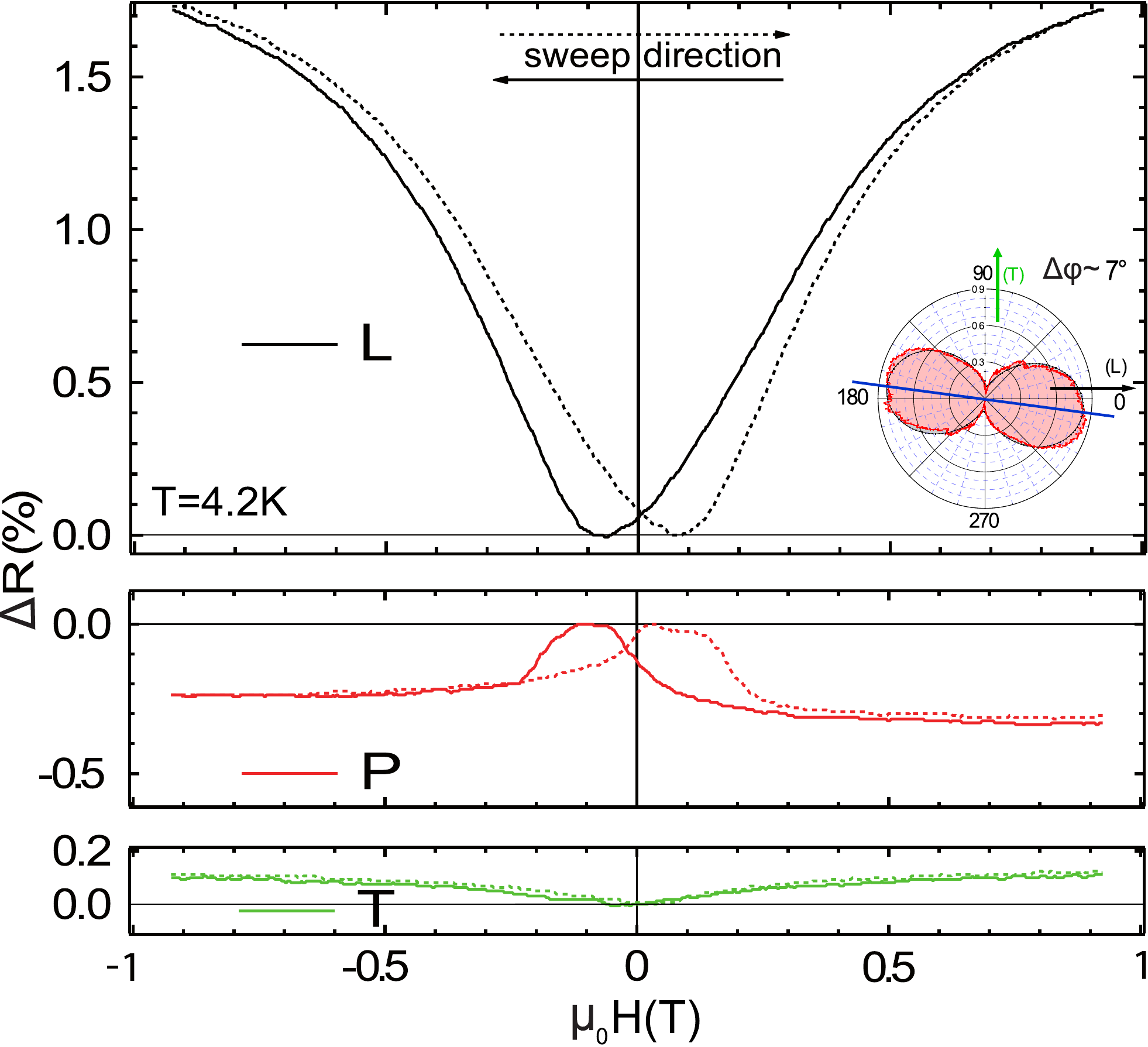}
\end{center}
\caption{(Color online) Anisotropic magnetoresistance $\Delta R = (R(\mu_0 H) -
R_{\un{max}}) / R_{\un{max}}$ of an individually contacted, $500\un{nm}$ wide
\PdNi\ strip. The three panels represent the relative orientiation of the
magnetic field according to Fig.~\ref{rem}, i.e. longitudinal (L, black),
perpendicular (P, red) and transverse (T, green). The dashed traces show the
magnetic field up-sweep, while the continuous traces show the down-sweep
measurement. All three panels have identical scaling of both axes. The inset shows a polar plot of the AMR at constant magnetic field $\left|\mu_0H\right| = 1\un{T}$ and reveals a small misalignment of $\Delta \phi \approx 7^{\circ}$
}
\label{amr}
\end{figure}
Fig.~\ref{amr} shows the relative change of resistance $\Delta R = (R(\mu_0 H) - R_{\un{max}}) /
R_{\un{max}}$, in dependence on the external magnetic field $\mu_0 H$, for a strip width of $500\un{nm}$.
The three panels correspond to the three field orientations as sketched in Fig.~\ref{rem}. The largest increase
of resistance occurs when applying the field in longitudinal (L) direction. This
means that a large part of the magnetization was oriented away from this
direction and had to be aligned parallel to the strip by the external magnetic field.
In perpendicular (P) direction we observe a decrease of the resistance with increasing perpendicular field, i.e. with the alignment of the magnetization perpendicular to the film plane and the in-plane current. 
This means that the magnetization must have a longitudinal component in the remanent state. 
In transverse (T) direction a small but finite increase of resistance appears with increasing field.
This is explained by a misalignment of the external field by $\un{7}^{\circ}$  relative to the (T) axis which was verified by measuring the angular variation of the magnetoresistance in a constant field of $\mu_0 H$ = 1 T (see inset of Fig. \ref{amr}).
A consistent interpretation of all three magnetoresistance curves in agreement with the 
SQUID magnetization curves is obtained by assuming that after saturation along a hard 
axis, i.e. either along an (L) or a (P) direction the magnetization is split into domains, as seen in Fig. \ref{mfm}(a), with
a small M component along (L) due to Néel type domain walls and close to the long edge of the strip.
The presence of domain walls in the remanent state is also responsible for the very similar hysteretic behaviour of the magnetization loops (Fig. \ref{squid}) and the magnetoresistance curves (Fig. \ref{amr}) along the (L) axis.  

The distinct reduction of the remanence along the T axis below 100\% due to the formation of some reversed domains (see Fig. \ref{mfm}(b)) has practiclly no effect on the magnetoresistance because the AMR scales with $\delta R \propto cos^2 (\alpha)$ while the magnetometer measures the parallel component of the average magnetization $\un{M}_{\parallel} \propto cos(\alpha)$ where $\alpha$ is the angle between the (T) axis and the domain magnetization.

As result we can conjecture that the spontaneous magnetization of the strips is
oriented mainly transverse to the the strips' long axis. The small amount of change in
resistance in a magnetic field along the transverse (T) direction means that the
majority of magnetic moments already was aligned in this direction before the
field was applied. The minima respectively maxima of the
magnetoresistance curves show hysteretic behaviour and are mirrored according
to the sweep direction of the external field.

\section{Magnetic Force Microscopy}

\begin{figure}[th]
\begin{center}
\includegraphics[width=8.5cm]{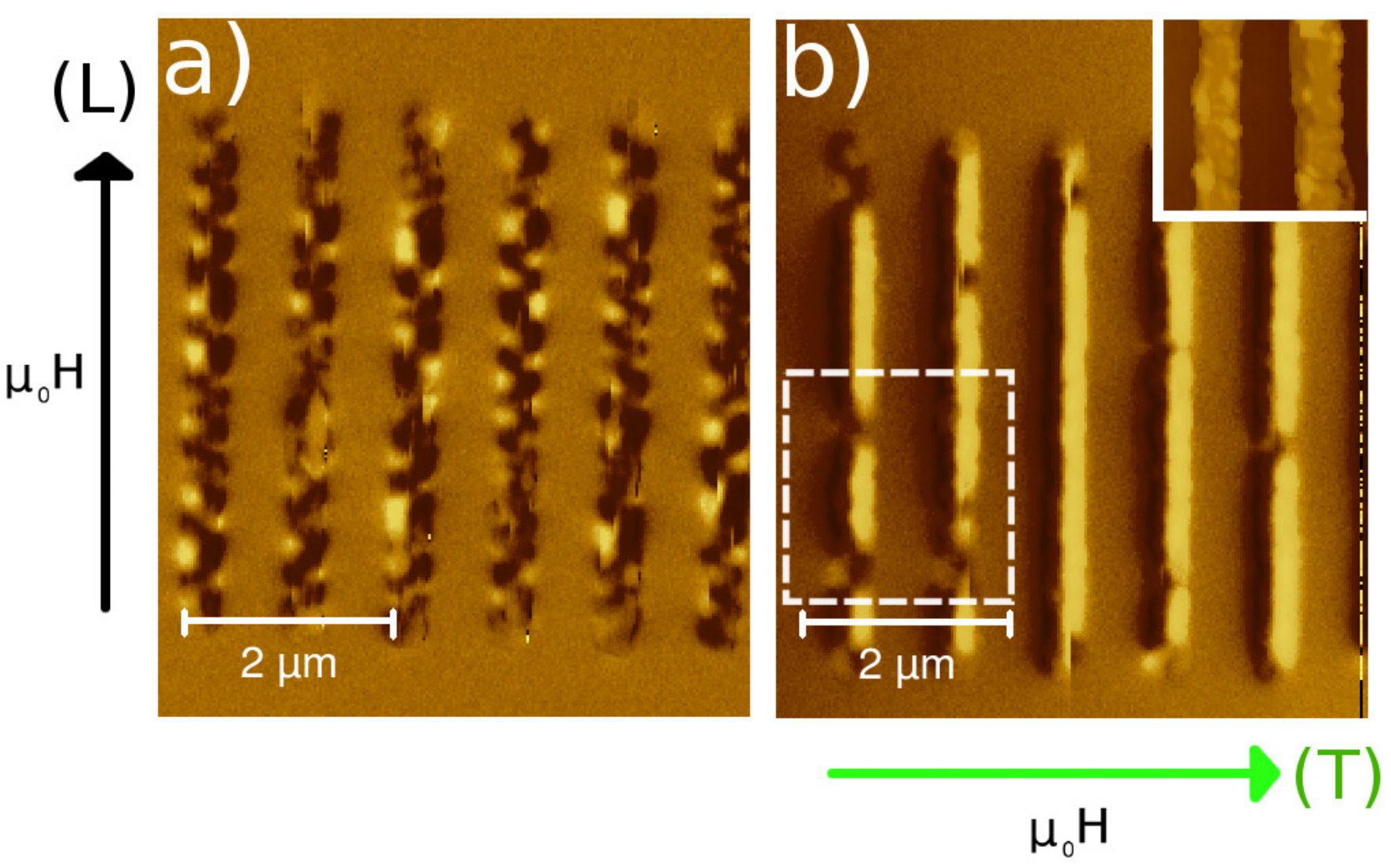}
\end{center}
\caption{(Color online) Magnetic force microscopy (MFM) images of the remanent
magnetization of an ensemble of 500nm wide \PdNi\ strips.
Before recording the image, the strips have been magnetized along the directions
indicated by the arrows, i.e. (a) in longitudinal (L), and (b) in transverse (T)
direction. The inset in (b) shows the surface topography of the region in the 
white dashed square}
\label{mfm}
\end{figure}

Finally, magnetic force microscopy (MFM) on a strip array directly reveals the
remanent micromagnetic configuration within narrow strips (width 500nm) of
\PdNi. Fig.~\ref{mfm} displays magnetic force microscopy (MFM) images of an
ensemble of contact strips. The MFM pictures were taken at room temperature and
zero external magnetic field; the scanning direction was longitudinal to the
strips (L). Prior to imaging, the sample was magnetized along the longitudinal
(L) (Fig.~\ref{mfm}(a)) or transverse (T) direction (Fig.~\ref{mfm}(b)) by an
external field of $\mu_{0}H=2\un{T}$ to ensure magnetic saturation.

The pattern in Fig.~\ref{mfm}(a), where the sample was previously magnetized in
longitudinal (L) direction, features alternating domain alignment along the
transverse (T) direction. This is consistent with a spontaneous magnetization in
transverse (T) axis, as inferred from the SQUID and AMR measurements.
This splitting in small domains in the remanent state also causes the non-zero remanence in the (L)
direction observed in the SQUID data (see Fig. \ref{squid}) as explained before.
In Fig.~\ref{mfm}(b), after magnetizing in (T) direction the magnetization structure in this direction is considerably more uniform. The strips feature large domains with transverse (T) magnetization
direction. The remaining disintegration into domains in certain areas may be
caused by, e.g.,  edge roughness of the strips, supported by thermal activated
processes at room temperature: in the inset of Fig.~\ref{mfm}(b), showing the
surface topography of the region marked by a white dashed square in the MFM picture,
several magnetic features re-appear as height profile irregularities.
Altogether, MFM imaging again confirms the findings of an easy axis along the
transverse (T) direction.

\section{Discussion}
A possible explanation for the transverse magnetic easy axis is the effect of
inverse magnetostriction \cite{chikazumi}. While many mechanisms can contribute
local stress during fabrication of the strips, it is already instructive to
look at the highly different thermal expansion coefficients $\alpha$ of the
thin film metal ($\alpha_{\text{Pd}} \approx 11.8 \times 10^{-6} \un{K}^{-1}$,
$\alpha_{\text{Ni}} \approx 13.4 \times 10^{-6} \un{K}^{-1}$) and the substrate
material used in this study ($\alpha_{\text{Si}} \approx 2.6 \times 10^{-6}
\un{K}^{-1}$). Considering a temperature higher than room temperature during
the thin film evaporation process of PdNi, tensile stress is imprinted on the
metal layer when the sample is cooled down. At the edges of the strips, this
stress can relax. If the strip is sufficiently narrow the relaxation of 
transversal stress takes place across the entire strip. Assuming a similar
behaviour of \PdNi\ and pure nickel (due to the high amount of nickel in our
alloy and the same crystal structure (fcc) of both elements), the magnetic
moments in the strip then align orthogonal to the remaining longitudinal stress,
i.e.\ in the transversal (T) direction, as the magnetostriction coefficient of
nickel is negative for both (100) and (111) direction\citep{chikazumi}.

\section{Conclusions}
As a conclusion, our work confirms in all aspects that the magnetic preferential
direction of high aspect ratio \PdNi\ thin film contact strips feature a magnetically easy axis transverse
to the strip orientation. This can be explained by the effect of
magneto-elastic coupling. It dominates other mechanisms as e.g.\ shape
anisotropy, which would favour a magnetically easy axis longitudinal to the
strip orientation. For
application of our investigated contact strips in CNT based spin devices it is
therefore recommended to apply external magnetic fields in transverse direction
in order to obtain a more distinct switching behaviour. Our results are well
compatible with those of the complementary study by Chaleau et al.
\citep{chauleau}, who did extensive micromagnetic simulations and investigated
the remanent magnetization direction using XMCD experiments.

\section{Acknowledgments}
The authors would like to thank J.-Y. Chauleau for helpful discussions. We
gratefully acknowledge funding by the EU FP7 project SE2ND and by the Deutsche
Forschungsgemeinschaft via SFB 689, GRK 1570, and Emmy Noether project Hu
1808/1-1.


\begin{thebibliography}{25}
\bibitem [{foo()}]{footnote1}%
  \BibitemOpen
  \href@noop {} {}\bibinfo {note} {{While local orbital moments may lead to an
  exception from this rule specifically in PdNi at Ni concentrations below
  50\%, this is not the case for the high Ni concentration present here (see
  Ref. \onlinecite{senoussi} for details).}}\BibitemShut {Stop}%
\end{thebibliography}
\end{document}